\documentclass[aps,prl,reprint,groupedaddress]{revtex4-1}

\usepackage{amssymb}
\usepackage{amsmath}
\usepackage{amsthm}
\usepackage{balance}
\usepackage{graphicx}
\usepackage{subfigure}

\usepackage{epstopdf}
\usepackage{amsfonts}
\usepackage{natbib}
\usepackage{hyperref}
\usepackage{color}
\usepackage{physics}

\usepackage{siunitx} 
\definecolor{midnightblue}{cmyk}{1,1,0,0.1}
\definecolor{forestgreen}{cmyk}{0.76,0,0.26,0.5}
\definecolor{cred}{RGB}{188,55,84}
\hypersetup{colorlinks=true,linkcolor=cred,citecolor=cred,urlcolor=cred}
\usepackage{bm}

\begin{document}

\title{ Valley-Contrasting Orbital Magnetic Moment Induced Negative Magnetoresistance}
\author{Hailong Zhou}
\email{zhou@utexas.edu}
\author{Cong Xiao}
\email{congxiao@utexas.edu}
\author{Qian Niu}
\affiliation{Department of Physics, The University of Texas at Austin, Austin, TX 78712, USA}

\date{\today }

\begin{abstract}
The valley-contrasting orbital magnetic moment of Bloch electrons allows the
lifting of valley degeneracy by an out-of-plane magnetic field. We
demonstrate that this leads to negative magnetoresistance, utilizing a gapped two-dimensional multi-valley model as
an example. An intuitive physical picture in terms of the increased carrier density from a magnetic gating effect is proposed for this negative magnetoresistance. In particular, giant negative magnetoresistance is
achieved after one of the two valleys is depleted by the magnetic field. This
new mechanism of negative magnetoresistance is argued to be relevant in
ionic-liquid gated gapped graphene with small effective mass.
\end{abstract}

\maketitle

The geometry of Bloch wave functions, especially the Berry curvature, has
profound implications on the electronic transport of crystalline solids \cite%
{XiaoReview, Weyl2018, TSMReview2018, TransportTSM2018}. The anomalous velocity of a Bloch electron resulted from the
Berry curvature leads to the anomalous Hall effect \cite{AHEreview,
DimiAHE2003, AHEExp} and valley Hall effect \cite{XiaoValley2007, Mak2014} in
time-reversal broken and space-inversion broken systems, respectively. In contrast to the substantially studied Berry curvature,
the importance of the orbital magnetic moment in inducing novel transport
behaviors has been noticed most recently. The orbital magnetic
moment can be pictorially thought of as arising from the self-rotation of a
wave-packet representing a semiclassical Bloch electron \cite{Chang1996},
and it couples to the magnetic field through a Zeeman-like interaction.
Therefore, it plays a central role in magneto-transport in the low-frequency
regime, such as the gyrotropic magnetic effect \cite{ZhongPRL2016, Pesin2015}
and dynamical magnetopiezoelectric effect \cite{VarjasPRL2016}.

The coupling between the orbital magnetic moment and the valley degree of
freedom in two-dimensional inversion-broken systems leads to the valley-contrasting orbital magnetic moment, which induces novel
phenomena \cite{Mak2018, Yao2016Review}, such as the valley-polarized zeroth
Landau-level in the high-field limit \cite{Cai2013, Li2013} and the
strain-induced valley magnetization \cite{Mak2017}. The valley-contrasting orbital magnetic moment enables the lifting of the valley degeneracy by an
out-of-plane magnetic field \cite{MacNeillPRL2015}, and yet how it affects the semiclassical magneto-transport has
not been fully revealed. Recently, Sekine and MacDonald addressed the
valley-dependent magnetoresistance \cite{Akihiko2018}. However, the role
played by the orbital magnetic moment was not discussed in their article.

This rapid communication aims to address the effect of the orbital magnetic moment, in addition to the Berry curvature, on transverse magnetoresistance in two-dimensional (2D) inversion-broken multi-valley systems. Based on the calculation for a 2D massive Dirac model \cite{XiaoValley2007} within the semiclassical transport theory, we predict a negative transverse magnetoresistance (orthogonal electric and magnetic fields, with the magnetic field perpendicular to the plane of the 2D system) emerges and strengthens with the increasing magnetic field. In relatively strong magnetic fields, we uncover that it is the orbital magnetic moment that dominates the behavior of magnetoresistance through valley-contrasting band-energy shift of Bloch electrons from Zeeman coupling, rather than the Berry curvature corrected phase space measure or the anomalous velocity. This valley-contrasting band-energy shift further causes a magnetic gating effect when it is associated with the energy-dependent density of states. Especially, a giant negative magnetoresistance appears from the efficacious magnetic gating effect when carriers in one of the two valleys are depleted by this band shift. At last, an ionic-liquid gating experiment involving gapped graphene with small effective mass such as graphene on hexagonal boron nitride (hBN-graphene) is discussed for possible observing of the predicted phenomenon.


\emph{{\color{blue} Preliminaries for semiclassical magneto-transport.}}%
---The semiclassical transport theory of Bloch electrons has three basic
ingredients, namely the semiclassical equations of motion, the phase-space
measure, and the occupation function which satisfies the semiclassical
Boltzmann equation.

The semiclassical equations of motion of a Bloch wave-packet are given by
\cite{XiaoReview}
\begin{subequations}
\label{eom}
\begin{eqnarray}
\dot{\bm{r}} &=&\tilde{\bm{v}}-\dot{\bm{k}}\times \bm{\Omega},
\label{eomrdot} \\
\hbar \dot{\bm{k}} &=&-e\bm{E}-e\dot{\bm{r}}\times \bm{B},
\end{eqnarray}
where $\tilde{\bm{v}}=\frac{1}{\hbar }\bm{\partial_{k}}\tilde{\varepsilon}$
is the group velocity modified by the Zeeman-like coupling between the
magnetic moment $\bm{m}$ and the weak external magnetic field through $\tilde{%
\varepsilon}=\varepsilon _{0}-\bm{B}\cdot \bm{m}$. Here $\varepsilon _{0}$
is the ordinary band energy, $\bm{m}$ generally consists of both the orbital
and spin magnetic moments.

On the other hand, the semiclassical measure for the number of quantum
states per unit volume in the phase-space is corrected by a factor
of $D_{\bm{k}}=1+\frac{e}{\hbar }\bm{B}\cdot \bm{\Omega}$ in systems with
non-zero Berry curvature $\bm{\Omega}$, to guarantee the conservation of
phase-space volume \cite{XiaoDoS2005}. With this correction, the occupation
function $f_{\bm{k}}$ of a wave-packet state, labeled by the crystal
momentum $\bm k$ (the band index is abbreviated), still complies with the
standard Boltzmann equation, which reduces to the following form for homogeneous
systems at steady state,
\end{subequations}
\begin{equation}
\frac{\partial f_{\bm{k}}}{\partial \bm{k}}\cdot \dot{\bm{k}}=-\frac{f_{%
\bm{k}}-f_{0}}{\tau }.  \label{beq}
\end{equation}%
The collision term is treated simply by relaxation time ($\tau $)
approximation, since we focus on effects of band geometric quantities. $%
f_{\bm{k}}=f_{0}+g_{\bm{k}}$, with $f_{0}$ and $g_{\bm{k}}$ the equilibrium
and non-equilibrium part of occupation function, respectively. Specifically, $f_{0}$ is defined as
\begin{equation}
f_{0}=f_{0}(\varepsilon _{0}-\bm{B}\cdot \bm{m}-\mu )=f_{0}(\tilde{%
\varepsilon}-\mu ).  \label{f0}
\end{equation}%

Note that in Eq. (\ref{f0}) a shift of $\delta \mu $ in chemical potential $\mu$ is possible in
the presence of the Berry-curvature \cite{XiaoDoS2005} and Zeeman-like coupling \cite{Gao2017}, if charge
neutrality is required, for instance, in three-dimensional (3D) topological systems. As for 2D systems, charge neutrality
is not always necessary since one can control the chemical potential within
a proper range by gating. Moreover, with ionic-liquid gating \cite{IonicLiquid2013, IonicLiquid2015}, a fixed chemical potential is practical in 2D multi-valley
systems with small band gaps. Justification of $\delta \mu =0$ is present in the
discussion section of this Rapid Communication.

In the semiclassical transport framework, the electric current is given by
\begin{equation}\label{current}
\bm{{J}}=-e\int [d\bm{k}]D_{\bm{k}}\dot{\bm{r}}f_{\bm{k}},
\end{equation}
with ${[d\bm{k}]}$ an abbreviation for
${d\bm{k}/}\left( {2\pi }\right) ^{2}$. In the linear response to
electric fields, one has $J_{i}=\sum_{j}\sigma _{ij}E_{j}$, where $i,j=x,y$
refer to the spatial indices and $\sigma _{xx}$ ($\sigma _{xy}$) refers to
the longitudinal (Hall) electrical conductivity. Once the
conductivity components are obtained, it is intuitive to illustrate the magnetic-field
effect on transport from the perspective of magnetoresistance, which is
defined by
\begin{equation}  \label{mre}
MR\equiv \frac{\rho _{xx}(B)-\rho _{xx}(0)}{\rho _{xx}(0)},
\end{equation}
if one recalls that the conventional result for the magnetoresistance of a
single-band system is zero, although the conductivities are $B$%
-dependent. This is so because the resistivity is converted from the
conductivity by
\begin{equation}  \label{rhoxx}
\rho _{xx}=\frac{\sigma _{xx}}{\sigma _{xx}^{2}+\sigma _{xy}^{2}},
\end{equation}
and $\rho _{xx} = \rho _{xx}(0)=1/\sigma _{xx}(0)$, provided with the relations $%
\sigma _{xy}=-\omega _{c}\tau \sigma _{xx}$ and $\sigma_{xx}=\sigma_{xx}(0)/(1+\omega _{c}^{2}\tau ^{2})$ for single band systems without
geometric effects involved. However, magnetoresistance is not zero any more
for systems with non-zero Berry curvature or magnetic moment, as we shall
demonstrate below.

\emph{{\color{blue} Qualitative picture: magnetic gating effect from valley-contrasting orbital magnetic moment.}}---A minimal model for 2D
inversion-broken multi-valley systems is the massive Dirac model adopted in \cite{XiaoValley2007, Akihiko2018}, which describes hBN-graphene \cite{Giovannetti2007} or graphene on SiC substrate \cite{ZhouSY2007} with Hamiltonian
\begin{equation}\label{model}
H=\hbar v_{F}(\tau _{z}k_{x}\sigma _{x}+k_{y}\sigma _{y})+h_{0}\sigma _{z},
\end{equation}%
where $2h_{0}$ is the band gap, $\sigma _{i}$ are Pauli matrices representing the sub-lattice. In particular, $\tau _{z}=\pm 1$ denotes $K$ and $K^{\prime }$ valley, respectively, indicating valley-contrasting physics \cite{XiaoValley2007}.

Noting that the Berry curvature and orbital magnetic moment in a 2D system are always in the normal ($z$) direction of the 2D ($xy$) plane, therefore, for conduction bands of model (\ref{model}) one has \cite{XiaoReview, XiaoValley2007}:
\begin{equation}\label{omegam}
\Omega _{z}= -\frac{\hbar ^{2}v_{F}^{2}h_{0}}{2\epsilon ^{3}}\tau_{z}, \text{ \ }m_{z}^{orb}=-\frac{e\hbar v_{F}^{2}h_{0}}{2\epsilon ^{2}}\tau_{z},
\end{equation}
with $\epsilon =\sqrt{( \hbar v_{F})^2(k_x ^2 + k_y^2)+h_{0}^{2}}$. For the same reason, the magnetic field $\bm{B} = B\hat{z}$ is applied along $z$ direction to enable the couplings of the Berry curvature and orbital magnetic moment with $\bm{B}$.

In systems with small band gaps and low Fermi level, the spin splitting in a magnetic field can be neglected, considering the orbital magnetic moment near the band edge is generally much greater than the spin moment \cite{XiaoValley2007}. Therefore, the Zeeman energy term contains only the orbital magnetic moment, leading to
\begin{equation}
\tilde{\varepsilon}=\varepsilon _{0}-B{m}_{z}^{orb}.  \label{Em}
\end{equation}

\begin{figure}[tbph]
\subfigure[]{\includegraphics[width=0.40\textwidth]{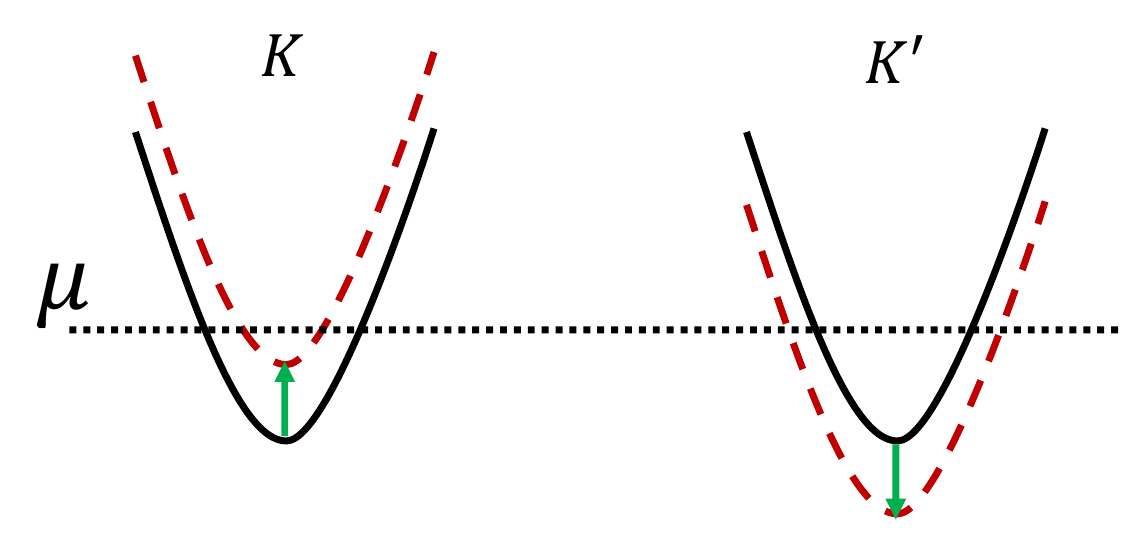}\label{illusta}}
\subfigure[]{\includegraphics[width=0.40\textwidth]{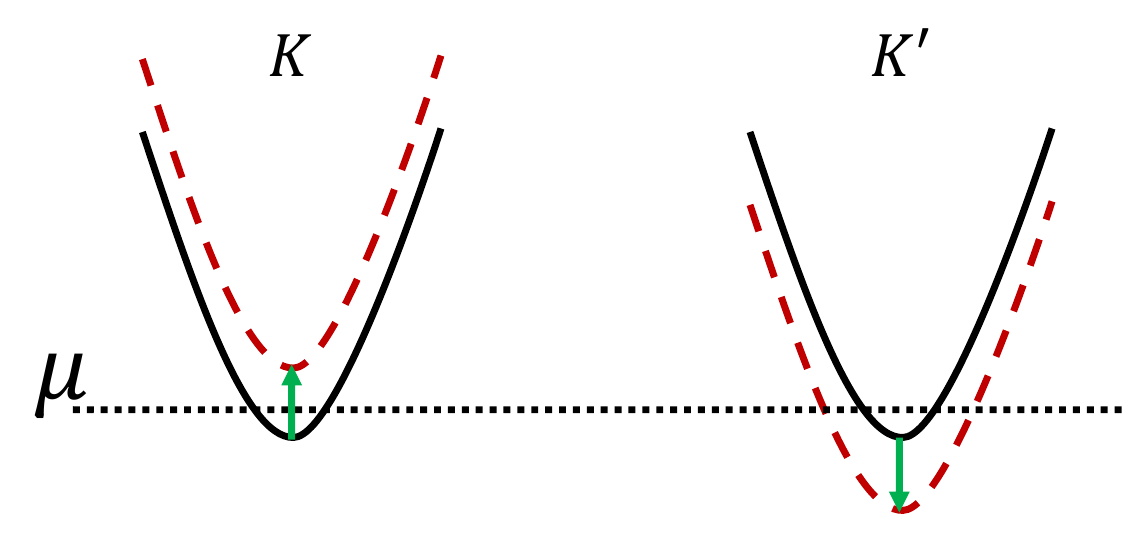}\label{illustb}}
\caption{(a) Chemical potential lies far from conduction band bottom. (b) Chemical potential sits closely to the conduction band bottom. In both (a) and (b), the solid (black) lines represent the band energy $\varepsilon_{0}$ without magnetic field, while the dashed (red) lines represent the orbital Zeeman energy corrected band energy $\tilde{\varepsilon}$, as shown in Eq. (\ref{Em}), for $K$ and $K^{\prime}$ valleys. In addition, the green arrows illustrate the directions of band shift due to orbital Zeeman energy, and the horizontal dotted (black) lines indicates where the chemical potential resides.}
\label{illust}
\end{figure}

The valley ($\tau_z$) dependence of the Berry curvature and orbital magnetic moment in Eq. (\ref{omegam}) gives rise to valley-contrasting effects on transport. For example, the correction term of the phase-space measure $\frac{e}{\hbar }B\Omega _{z}$ for $K$ and $K'$ valley carries different signs, suggesting that the response to a magnetic field for the two valleys is also opposite in sign. More importantly, due to the valley-contrasting orbital magnetic moment, the orbital Zeeman energy $-Bm_{z}^{orb}$ shifts energy bands of the two valleys to the opposite directions, i.e. if the energy band of $K$ valley shifts up, then the energy band of $K^{\prime }$ valley shifts down, and vice versa, resulting in valley-contrasting band energy shift, as shown in Fig. \ref{illust}.

This valley-contrasting energy shift induces a so-called magnetic gating effect by increasing the total carrier density in the system, noticing the asymmetry of the density of states of $K$ and $K^{\prime }$ valleys at the Fermi level in the presence of a finite magnetic field. As illustrated in Fig. \ref{illust}, when $B$ is positive, the density of states at the Fermi level in $K^{\prime }$ valley $D_{K^{\prime}}\left( \mu \right) $ is always larger than that in $K$ valley $D_{K}\left(\mu \right) $ (specially, in the case of Fig. \ref{illustb}, $D_{K}\left( \mu \right) =0$), since the density of states increases with energies. Thus, the change of the carrier density is approximately proportional to
\begin{equation}\label{carrierdensity}
\delta n_{e}\sim \left[ D_{K^{\prime }}\left( \mu \right) -D_{K}\left( \mu\right) \right] {\chi }B>0,
\end{equation}
where $m_{z}^{orb}=-\chi \tau _{z}$. This $\delta n_{e}$ is positive definite, indicating that the total carrier density increases with the magnetic field, resulting in a magnetic gating effect.

From the increasing carrier density, we can anticipate the negative magnetoresistance immediately, given that the resistance is inversely proportional to the carrier density. Furthermore, because the orbital magnetic moment of the conduction band is mostly concentrated around the band bottom, the negative magnetoresistance is expected to be strong when the Fermi level lies closer to the band bottom. More dramatically, after carriers in one of the two valleys are completely depleted, as illustrated in Fig. \ref{illustb}, corresponding to $D_K(\mu) = 0$ in Eq. (\ref{carrierdensity}), the the resistance reduction is much more efficient since the carrier density increases with higher efficiency than that prior to depletion.

\emph{{\color{blue} Quantitative results from Boltzmann calculation.}}---Now we support the above physical picture with quantitative analysis. In 2D systems with the magnetic field along $z$ direction, equations of motion (\ref{eom}) lead to
\begin{subequations}\label{rdotkdot}
\begin{eqnarray}
D_{k}\dot{\bm{r}} &=&\tilde{\bm{v}}+\frac{e}{\hbar }\bm{E}\times \bm{\Omega},
\label{rdot} \\
D_{k}\dot{\bm{k}} &=&-\frac{e}{\hbar }[\bm{E}+\tilde{\bm{v}}\times \bm{B}].
\label{kdot}
\end{eqnarray}
\end{subequations}

It is worth noting that we do not expand $\dot{\bm{r}}$ and $\dot{\bm{k}}$ in Eqs. (\ref{rdotkdot}) in terms of the magnetic field, because the magnetic field enters these equations through several mechanisms, such as orbital Zeeman energy, phase-space measure, anomalous velocity, and Lorentz force. These mechanisms may be governed by distinct characteristic magnetic fields with various magnitudes, thus, a given magnetic field may be weak for one mechanism but strong for another. Therefore, it is challenging to ensure the validity of a simple expansion with respect to the magnetic field prior to understanding these characteristic scales. In fact, the semiclassical magneto-transport theory \textit{without} the aforementioned expansion agrees well with the experimental data for the longitudinal magnetoresistance 3D topological insulators in a wide regime of magnetic fields \cite{Dai2017}, while the theory \textit{with} the aforementioned expansion failed in accomplishing comparable agreement \footnote{Z. Z. Du, private communication.}.

With the Eq. (\ref{kdot}), the semiclassical Boltzmann equation (\ref{beq}) can be simplified to the
form below:
\begin{equation}
-\frac{1}{\hbar D_{k}}\frac{\partial f_{0}}{\partial \bm{k}}\cdot e\bm{E}-%
\frac{1}{\hbar D_{k}}(e{\tilde{\bm{v}}}\times \bm{B})\cdot \frac{\partial g_{%
\bm{k}}}{\partial \bm{k}}=-\frac{g_{\bm{k}}}{\tau }.  \label{simbeq}
\end{equation}%
Considering the case with negligible intervalley scattering (this is reasonable in gapped graphene of high electronic quality \cite{Geim2014}), it is reasonable to simply solve the Boltzmann equation for each valley separately in a 2D
isotropic system, where the following ansatz for $g_{\bm{k}}$
\begin{equation}
g_{\bm{k}}=(-\frac{\partial f_{0}}{\partial \tilde{\varepsilon}})(-\tau
^{\parallel }(\bm{B},\tilde{\varepsilon})e\bm{E}-\tau ^{\perp }(\bm{B},%
\tilde{\varepsilon})\hat{z}\times e\bm{E})\cdot \tilde{\bm{v}}
\label{ansatz}
\end{equation}%
in linear response to the electric field can be employed.

Plugging Eq. (\ref{ansatz}) into Eq. (\ref{simbeq}) yields:
\begin{subequations}
\begin{eqnarray}
\tau ^{\parallel } &=&\frac{\tilde{\tau}}{1+{\tilde{\omega}_{c}^{2}\tilde{%
\tau}^{2}}}, \\
\tau ^{\perp } &=&\frac{{\tilde{\omega}_{c}\tilde{\tau}}}{1+\tilde{\omega}%
_{c}^{2}\tilde{\tau}^{2}}\tilde{\tau}=\tilde{\omega}_{c}\tilde{\tau}\tau
^{\parallel },
\end{eqnarray}%
with $\tilde{\tau}=\tau /D_{k}$ and $\tilde{\omega}_{c}={eB}/\tilde{m}$.
Here $\tilde{m}(\tilde{\varepsilon})\tilde{\bm{v}}=\hbar \bm{k}$, where $%
\tilde{m}$\ is a magnetic field dependent (through $\tilde{\varepsilon}$)
quantity with the dimension of mass. Here we point out that the solution of $g_{\bm{k}}$ reduces to the result of conventional Hall effect \cite{Ziman} immediately in systems with neither Berry curvature nor magnetic moment.

Then, the electrical conductivities are obtained from Eq. (\ref{current}) as
\end{subequations}
\begin{equation}
\sigma _{xx}=e^{2}\int {[d\bm{k}]\tilde{v}_{x}^{2}(-\frac{\partial f_{0}}{%
\partial \tilde{\varepsilon}})\tau ^{\parallel }},  \label{condxx}
\end{equation}%
and
\begin{equation}
\sigma _{xy}=-\frac{e^{2}}{\hbar }\int {[d\bm{k}]\Omega _{z}f_{0}}-e^{2}\int
{[d\bm{k}]\tau ^{\perp }\tilde{v}_{y}^{2}(-\frac{\partial f_{0}}{\partial
\tilde{\varepsilon}})}.
\label{condxy}
\end{equation}%
The second term of $\sigma _{xy}$ resembles the ordinary Hall conductivity from Lorentz force, but with corrections from the Berry curvature and orbital moment. The first term in Eq. (\ref{condxy}) is the integral of the Berry curvature over occupied states. This term does not contribute to the total Hall conductivity in the presence of time-reversal symmetry if one adds the contributions from the two valleys together \cite{XiaoValley2007}, but can be nonzero here because the valley-contrasting energy shift due to the orbital magnetic moment (Eq. (\ref{f0})) breaks the time-reversal symmetry explicitly.

To evaluate the behavior of magnetoresistance, we numerically calculate $\sigma _{xx}$ and $\sigma _{xy}$ for each valley. Then we add contributions from two valleys together to get the full conductivity components. The longitudinal resistivity $\rho _{xx}$ and transverse magnetoresistance are calculated from Eq. (\ref{rhoxx}) and Eq. (\ref{mre}).

\begin{figure}[ptbh]
\subfigure[]{\includegraphics[width=0.23\textwidth]{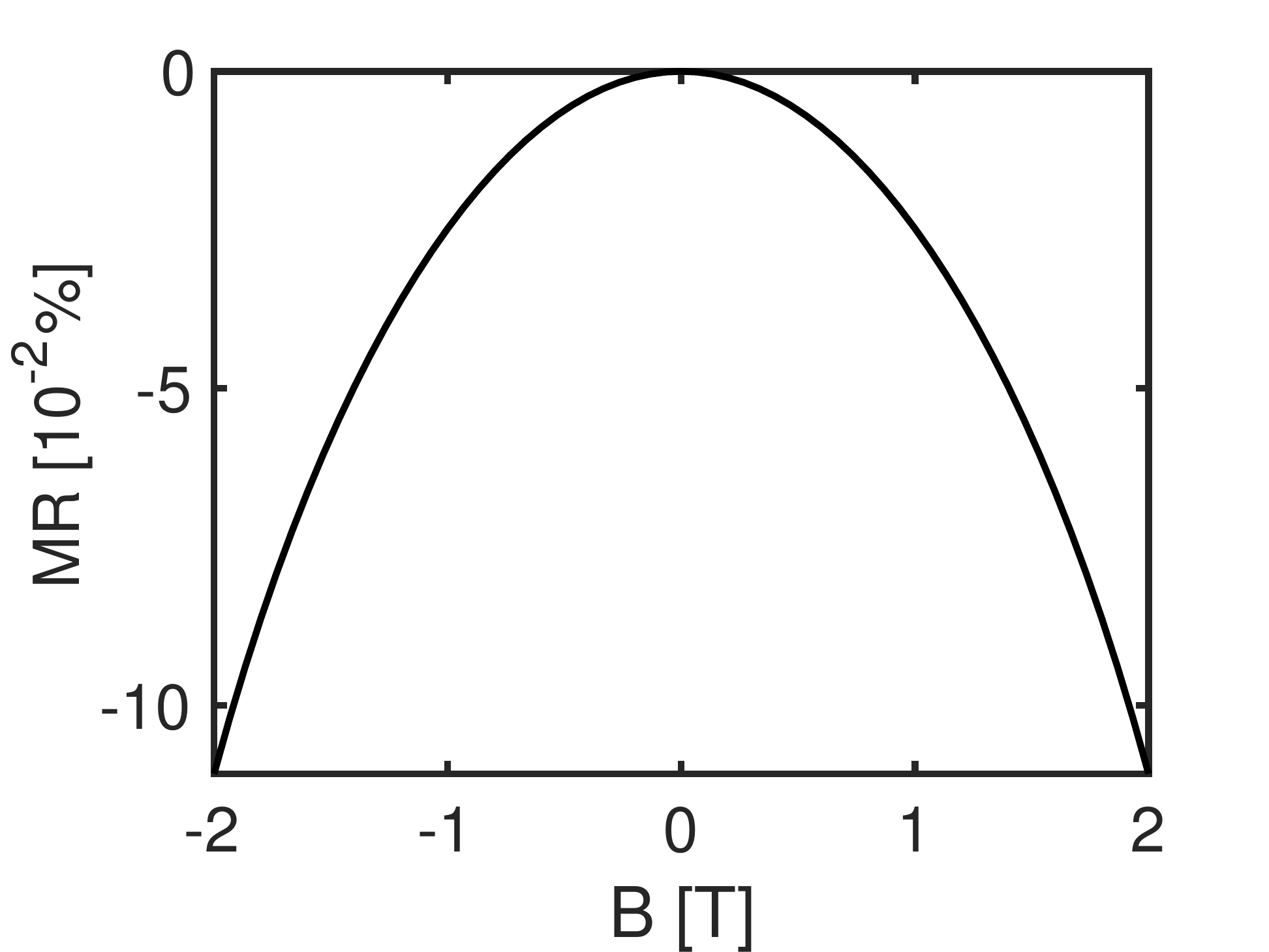}\label{36mev}}
\subfigure[]{\includegraphics[width=0.23\textwidth]{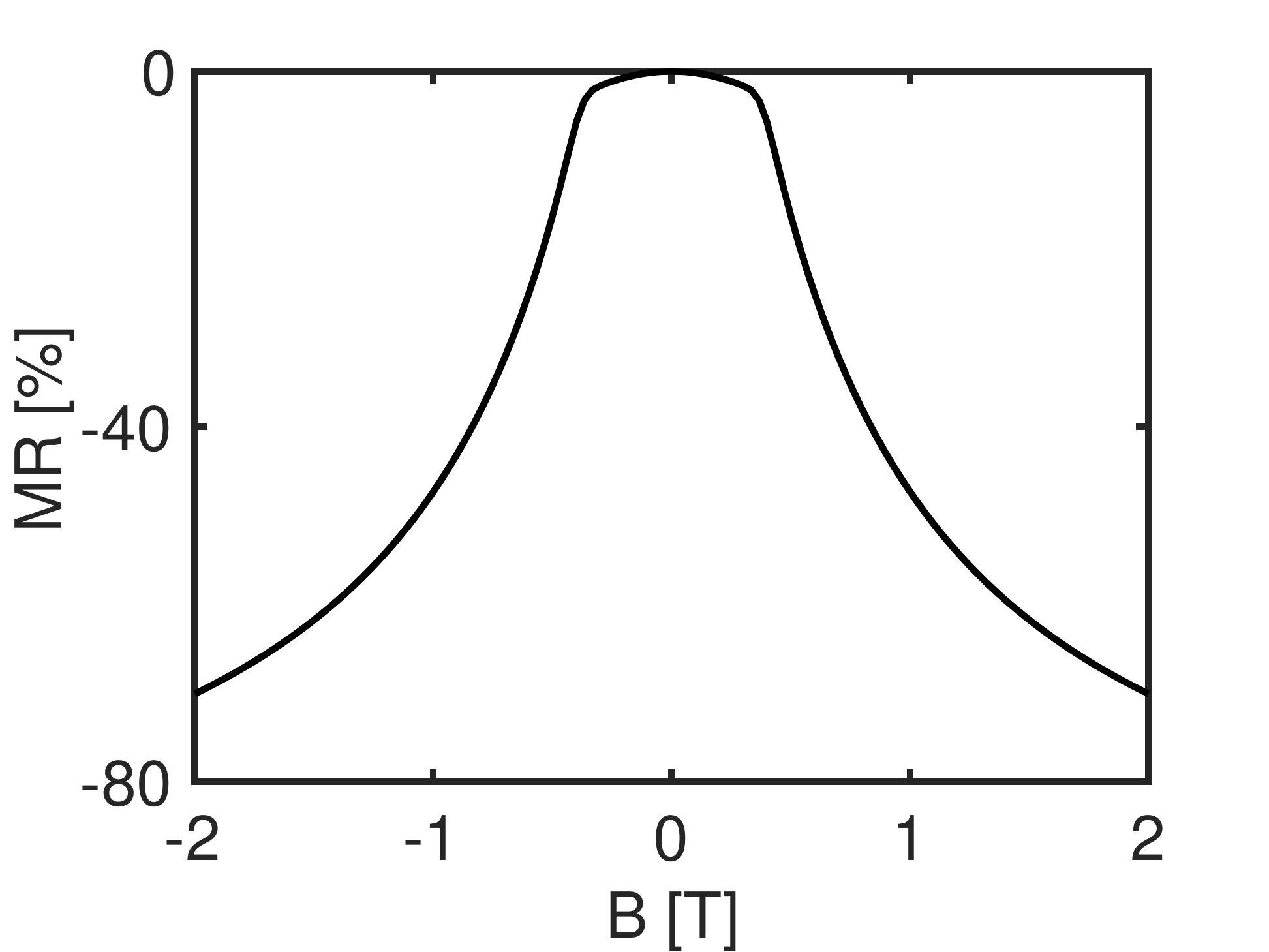}\label{27mev}}
\caption{(a) MR varies with the magnetic field for case (1) $\mu = 36\ \si{m.\electronvolt}$. (b) MR for case (2) $\mu = 27\ \si{m.\electronvolt}$. In both (a) and (b), the parameters are chosen as $h_0 = 26\ \si{m.\electronvolt}$, $\hbar v_F = 3.0\ \si{\electronvolt}\si{\angstrom}$, $\tau = 10^{-13}\ \si{s}$, $T = 1.5\ \si{K}$.}
\label{mrf}
\end{figure}

Given the discussions above, we adopt $h_0 = 26\ \si{m.\electronvolt}$ to achieve a band gap of $52 \ \si{m.\electronvolt}$ for hBN-graphene \cite{Giovannetti2007} and compare the magnetoresistance for two different cases: (1) Fermi level lies relatively far from the band bottom, as shown in Fig. \ref{illusta}; (2) Fermi level lies closely to the band bottom, as shown in Fig. \ref{illustb}. For case (1), the negative magnetoresistance from geometric effects is vanishingly small as expected, about $-0.11\%$ at $B=2\ \si{T}$, as shown in Fig. \ref{36mev}, when $\mu=36\ \si{m.\electronvolt}$. For case (2), the geometric effects on magnetoresistance are significant: MR reaches about $-70\%$ at $B=2\ \si{T}$ from Fig. \ref{27mev}, when $\mu =27\ \si{m.\electronvolt}$.

This giant negative magnetoresistance results from the efficient magnetic gating effect after carriers in one valley is completely depleted by the magnetic field through orbital Zeeman shift. The critical magnetic field $B_{c}$ required to deplete carriers in one valley can be estimated by $B_{c}m_{z}^{orb}=\mu -h_{0}$, and it locates at the turning point from a rather flat quadratic-like curve to a more dramatic decreasing rate of $1/B$ in Fig. \ref{27mev}.

In relatively weak magnetic fields, that is when $B$ is smaller than $B_c$ mentioned above, the negative magnetoresistance is attributed to the non-zero Berry curvature in phase space measure, orbital moment corrections to energy and group velocity, and the entanglement of the Berry curvature and orbital magnetic moment in the anomalous Hall term in Eq. (\ref{condxy}). Among these contributions, neither Berry curvature nor orbital magnetic moment dominates the total negative magnetoresistance. In this case, one could summarize these contributions as geometric contribution and expand the conductivity according to the order of the magnetic field, as shown in \cite{Gao2017, Chenhua2015}. However, in relatively strong magnetic fields, specifically when $B$ is greater than $B_c$, it is ambitious to justify an expansion according to the order of the magnetic field, as we discussed previously. Fortunately, in this regime, we are able to identify (by virtually turning on only either the Berry curvature or orbital magnetic moment at one time in numerical calculations) that it is the orbital magnetic moment that dominates the negative magnetoresistance \footnote{Supplemental Material}, through the valley contrasting band shift and magnetic gating effect.

\emph{{\color{blue}Summary and discussion.}}---In summary, we demonstrate that the valley-contrasting orbital magnetic moment induces giant negative magnetoresistance in spacial-inversion broken multi-valley systems by a magnetic gating effect, emerging from the combination of the valley contrasting band shift and the energy dependent density of states. According to our knowledge, this mechanism for negative magnetoresistance has not been noticed in previous researches.

This new mechanism of negative magnetoresistance is a result of valley-contrasting magnetic gating effect, which essentially provides more carriers as the magnetic field increases while keeping the chemical potential unchanged. Although it seems challenging to satisfy $\delta \mu = 0$ for many experimental setups, it is certainly practical in the context of ionic-liquid gating, where the effective capacitance at ionic-liquid gate can be as high as $C_{ILG} = 7.8 \times 10^{-6}\ \si{F. cm^{-2}}$, estimated from the reported ionic-liquid gate coupling efficiency for a 2D few-layer $\text{MoS}_{\text{2}}$ transistor \cite{IonicLiquid2015}. The quantum capacitance of the 2D material we study here is $C_Q = g_v m^{*}e^2/\pi\hbar^2 \approx 6.1\times 10^{-7}\ \si{F. cm^{-2}}$ with effective mass $m^{*} = 4.7\times10^{-3}m_e$ for a gapped graphene model with band gap $53\ \si{m. \electronvolt}$ \cite{Giovannetti2007}. In this case, with $C_Q \ll C_{ILG}$, the effective band shift $\delta E$ induced by the orbital Zeeman energy $\delta E_B = -Bm^{orb}_z$ is given by $\delta E = \delta E_B/(1 + C_Q/C_{ILG}) \approx \delta E_B$, guaranteeing $\delta \mu = \delta E - \delta E_B \approx 0$ as stated previously. Thus, ionic-liquid gated gapped graphene with small effective mass is a compelling candidate for realizing the predicted giant negative magnetoresistance. In systems that do not satisfy the condition $C_Q \ll C_{gate}$, e.g. systems with larger effective mass (larger quantum capacitance) or systems with smaller gating capacitance $C_{gate}$, the chemical potential will not stay fixed. The change in chemical potential compensates the energy band shift from orbital Zeeman energy, resulting in a weaker magnetic gating effect, thereby weakening the negative magnetoresistance.

Lastly, we discuss the validity of the relaxation time approximation, a simplified treatment regarding the disorder scattering adopted during solving the Boltzmann equation. As is well known from studies on the anomalous Hall effect and orbital magnetoelectric response \cite{AHEreview, Pesin2017}, this simplification includes the key geometric effects while neglecting two accompanying delicate scattering processes: the side-jump and skew scattering. However, neglecting these two scattering effects does not appear to curtail the excellent agreements between experimental results and theoretical predictions based on merely geometric mechanisms in systems that the Fermi level locates near the band edges, as shown by measurements of valley Hall effect in hBN-graphene \cite{Geim2014}. Thus, in our study of magnetoresistance, the relaxation time approximation, which captures the geometric contributions, is adequate to comprehend the essential physics.

\begin{acknowledgments}
We thank H. Chen, B. Xiong, Y. Gao for helpful discussions. Q.N. is supported by DOE (DE-FG03-02ER45958, Division of Materials Science and Engineering) on the model analysis in this work. H.Z and C.X. is supported by NSF (EFMA-1641101) and Welch Foundation (F-1255).
\end{acknowledgments}

\balance
\bibliographystyle{apsrev4-1}

\end{document}